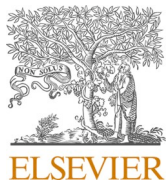
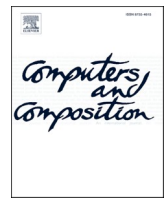

# "Wayfinding" through the AI wilderness: Mapping rhetorics of ChatGPT prompt writing on X (formerly Twitter) to promote critical AI literacies

Anuj Gupta [*], Ann Shivers-McNair

*University of Arizona, Department of English, Tucson, AZ 85712, United States*

ABSTRACT

In this paper, we demonstrate how studying the rhetorics of ChatGPT prompt writing on social media can promote critical AI literacies. Prompt writing is the process of writing instructions for generative AI tools like ChatGPT to elicit desired outputs and there has been an upsurge of conversations about it on social media. To study this rhetorical activity, we build on four overlapping traditions of digital writing research in computers and composition that inform how we frame literacies, how we study social media rhetorics, how we engage iteratively and reflexively with methodologies and technologies, and how we blend computational methods with qualitative methods. Drawing on these four traditions, our paper shows our iterative research process through which we gathered and analyzed a dataset of 32,000 posts (formerly known as tweets) from X (formerly Twitter) about prompt writing posted between November 2022 to May 2023. We present five themes about these emerging AI literacy practices: (1) areas of communication impacted by prompt writing, (2) micro-literacy resources shared for prompt writing, (3) market rhetoric shaping prompt writing, (4) rhetorical characteristics of prompts, and (5) definitions of prompt writing. In discussing these themes and our methodologies, we highlight takeaways for digital writing teachers and researchers who are teaching and analyzing critical AI literacies.

## 1. Introduction

Long before and during the increased uptakes of generative AI in late 2022 and early 2023, scholars in writing studies have engaged with the possibilities and pitfalls of automated writing, including automated writing evaluation systems in the classroom (Huang & Wilson, 2021), automated writing systems generating political content online (Laquintano & Vee, 2017), automated accessibility content like alt text (Zdenek, 2020), automated content management technologies (Andersen, 2007), automated data collection and analysis techniques (Gallagher, 2019), and automated content produced by chatbots (Ding et al., 2019). A closely related thread of scholarship engages with the rhetorical impacts of machine learning algorithms on human-generated writing practices, particularly in social media (e.g., Adams et al., 2020, Gallagher et al., 2020; Glotfelter, 2019, Koenig, 2020, Rice, 2020, Shepherd, 2020). And, with the rise of generative AI, scholars have called our field to engage critically and carefully with AI in our research and practice (e.g., Byrd, 2023; Graham & Hopkins, 2022, Duin & Pedersen, 2023; Majdik and Wynn 2023) and in our teaching (e.g., Gonzales, 2023, Gupta et al., 2024, Gupta, 2024a, 2024b, Tham et al., 2022).

* Corresponding author.
 *E-mail addresses:* anujgupta@arizona.edu (A. Gupta), shiversmcnair@arizona.edu (A. Shivers-McNair).

In computers and writing, specifically, scholars have long attended to the need for critical engagement with AI. As Ranade and Eyman (2024) point out in their introduction to a recent special issue of this journal on composing with generative AI, the first issue of *Computers and Composition* addressed the potential of artificial intelligence in the writing classroom (Burns 1983), and by 2018, scholars like Hart-Davidson (2018) were predicting that AI would take over early drafting in the writing process, with humans stepping in at the revision stage. Ranade and Eyman (2024) likewise recognize potential for "increased productivity," "improved quality" of writing, "expanded possibilities" and ideas, and the "democratization of creativity," even as they also highlight the risks of "bias" embedded in generative AI tools, the "lack of transparency" about how the tools work, and "security risks" including the creation of "malicious content" (p. 2).

Our study contributes to this body of work by analyzing emerging social media rhetorics of ChatGPT as a means to prompt critical AI literacies. We began this study when, in late 2022 and early 2023, we noticed a huge upsurge in public conversations about generative AI on social media. As researchers who are also teachers, we recognized a need to understand public discourses about writing with and for ChatGPT as a way of paying "closer attention to the resources" students are engaging with that "allow them to be so flexible" in moving among writing contexts, as Alexander et al. (2020) call for (p. 124). Critically examining them would help us prepare ourselves to teach students to engage critically and skillfully with generative AI. Like Ranade and Eyman (2024) and the authors in their special issue, we sought to "make sense of the affordances and constraints of these tools" so that we and our students could "use them effectively for writing as communication, writing as learning, and writing as thinking" (p. 3). In addition to our own interactions with generative AI tools, we sought to understand circulating discourses about the affordances and constraints of generative AI tools; therefore, we turned to social media data to help us identify and examine emerging rhetorics about writing with and for tools like ChatGPT.

We specifically turned to the social media platform X (formerly Twitter)[1] due to its long established credibility as an important reflection of various kinds of public discourses in social media research in writing studies (Buck, 2023; Coad, 2017; Dighton, 2020; Graham, 2021; Holmes & Lussos, 2018; Lussos, 2018; Walker & Laughter, 2019; Wolff, 2018). In a study that was approved by our Institutional Review Board (IRB), we obtained a dataset of over 250,000 posts (formerly known as tweets)[2] from November 2022 to May 2023 that contained keywords "ChatGPT" and "writing." In an iterative analysis process that blended descriptive statistics, close-reading, and topic modeling, we narrowed our focus to examine a subset of approximately 32,000 posts about "prompts," a term which has recently started referring the instructions users write to tools like ChatGPT. In that smaller dataset focused on prompts, we used qualitative coding to identify five themes in emerging rhetorics of prompt writing[3], and we analyzed those themes for implications for critical writing, rhetoric, and technical communication pedagogy and for further research. This decision to focus our analysis on prompts and prompt writing in our dataset happened in concomitance with emerging advisory guidelines by professional organizations in writing studies like the MLA-CCCC Joint Task Force on Writing and AI about the vital need to teach students how to write ChatGPT prompts as a means to develop critical AI literacies (MLA-CCCC Joint Task Force, 2023).

In what follows, we first situate our approach within traditions of digital writing scholarship that we draw on. Next, we describe our positionalities and reflect on how they impact our study. Next, we present our iterative analysis process that led us to the focus on ChatGPT prompts and prompt writing that we highlight in this article. As we present our methods, we acknowledge that the social media research landscape has changed since we undertook our study—something that is potentially always happening due to the past faced nature of innovations in digital rhetoric technologies. Thus we also underscore the need for teacher-researchers seeking to engage with emerging digital rhetorics to critically inform their pedagogies to develop flexible and agile approaches. We do this by describing an iterative, interactive methodology through which we arrived at a focus on prompt writing and, ultimately, the analysis of emergent rhetorics that, we argue, can inform how we teach and understand prompt writing. We then go onto present our findings, where we discovered five emerging themes about these emerging AI literacy practices on X: (1) areas of communication impacted by prompt writing, (2) micro-literacy resources shared for prompt writing, (3) market rhetoric shaping prompt writing, (4) rhetorical characteristics of prompts, and (5) definitions of prompt writing. In presenting and discussing themes from our analysis as well as our methodologies, we highlight takeaways for digital writing teachers and researchers who are teaching and analyzing critical AI literacies in their contexts. Our work also models values and practices that we recommend are helpful for engaging with dynamic,

---

[1] While our research was conducted at a time when the platform was still called "Twitter", since we are publishing the paper at a time when it has transitioned to being called "X", we have respected the platform's new guidelines and referred to it as "X" in our paper wherever possible. However, in instances when we are quoting or paraphrasing from research that was carried out before this transition happened, we have retained the term "Twitter" as originally used by the authors. We have also used the term "Twitter" when referencing features representative primarily of the old platform. Additionally, when discussing something that relates to both older and newer versions of the platform, we have used the phrase "X (formerly Twitter)" to index both versions.

[2] While this research was conducted at a time when texts on the platform were still called "tweets," we have respected the platform's new guidelines and referred to them as "posts" in our paper wherever possible. However, when quoting or paraphrasing from research that was carried out much before this transition happened, we have retained the term "tweets" as used by the original authors. We have also used the term "tweets" when referencing features representative primarily of the old platform.

[3] We are using the term "prompt writing" instead of "prompt engineering" to refer to the processes of composing prompts in response to Selber (2024). He argued that the phrase "prompt engineering" furthers the appropriation of GenAI prompts by a discourse of engineering which has brought with it a set of techno-utopic values. Such discourse presents prompts as "silver bullets" that can magically automate all writing with minimal human intervention. He advises writing and rhetoric scholars to reclaim this term and bring our disciplinary expertise to help shape its understandings in the public that help us realise that prompting is an iterative, rhetorical process that requires human expertise, critical thinking, and tinkering.

disruptive, and emergent technologies in research and teaching.

**2. Digital writing traditions we draw on**

Our decisions to engage with social media rhetorics about ChatGPT and to use machine learning approaches in our analysis were informed by traditions of studying digital writing that have modeled contextual, critical, descriptive and innovative approaches for engaging with dynamic writing technologies and practices. These traditions include how we frame literacies, how we study social media rhetorics, how we engage iteratively and reflexively with methodologies and technologies, and how we blend computational methods with qualitative methods. We briefly describe below how these traditions have informed our thinking and our research choices.

*2.1. 'Wayfinding' orientation to literacy*

We begin by sharing a terministic screen, to use Burke's (1966) metaphor for the ways that terminology "directs the attention" (p. 45), through which we orient to digital literacies. Our goal is to take a descriptive, rather than prescriptive, approach to understanding emergent digital literacies, following long standing approaches like those of the Building Digital Literacy research cluster of the Digital Life Institute, who discourage orienting to digital literacy "from a prescriptive approach that focuses mainly on productivity and proficiency," but, instead, emphasize the "need not just to assign competency-based heuristics but also to strive to understand what's happening when students engage and use new digital technologies" (Tham et al., 2021, p. 5). To this end, we found writing studies researchers Alexander, Lunsford, and Whithaus' concept of "wayfinding" (Alexander et al., 2020; Lunsford et al., 2024; Whithaus et al., 2022) a fruitful way of orienting to the complexity of emerging digital literacies.

Wayfinding, as a term and concept, has multiple uptakes in writing studies. For example, online writing leadership scholars Harris and Thomas (2023) draw on a definition from interface design (via Lynch & Horton, 2016) to emphasize wayfinding as a "decision-making process" that includes "four cognitive actions: orientation, route decisions, mental mapping, and closure" (p. 304), and they apply this definition to the design of online writing environments that support learners. Though there are distinct differences, this uptake of wayfinding shares some common ground with Alexander, Lunsford, and Whithaus' conceptualization of wayfinding, who similarly acknowledge that wayfinding is a technical term in a wide range of fields (Alexander et al., 2020, p. 121) that draws attention to people's intentional decisions as they navigate a physical or virtual space.

But the group's conceptualization of wayfinding also includes attention to the ways in which environments and interacting agents (both human and nonhuman) exert multi-directional influence on a person's wayfinding (Lunsford et al., 2024, p. 273). Relatedly, the group also positions their conceptualization of wayfinding as an alternative to more unidirectional models of accounting for writing development, like transfer, that theorize domains and knowledges as more-or-less discrete (Whithaus et al., 2022, p. 2). Their initial presentation of the concept, in Alexander et al. (2020), both builds upon and distinguishes itself from a genealogy of literacy metaphors in writing studies that orient to writers' practices across contexts, including worlds apart (e.g., Dias et al. 2013), literacy in the wild (e. g., Alvarez et al. 2017), ecologies and networks (e.g., Spinuzzi 2008, 2015), and transfer (e.g., Moore and Bass 2017). Key to the group's conceptualization of wayfinding is an attunement to a "continual process that is far from linear and may be significantly more recursive than our current theoretical models account for" (Whithaus et al., 2022, p. 2).

It is this attunement to complexity that particularly compels us in the group's conceptualization of wayfinding as a way of orienting to literacy that is attuned to how writers navigate multidimensional, fluid, constantly evolving rhetorical landscapes and writing technologies across personal, academic, professional, and public domains. Alexander et al. also build upon computers and composition scholarship (e.g., Selfe and Hawisher 2004) that highlights how emerging and evolving technologies mediate (and are mediated by) literacy practices across contexts. Thus, Alexander et al. (2020) define wayfinding as "a searching, but also a doing–a working, creating, and discovering process" (p. 122) as people navigate "the rise of new technologies and media [that] has led to the creation of new communities and forms of expression with quickly evolving conventions that mediate the lived reality" of writers (p. 124). We find this attunement to dynamic and complex environments, with attention to the ways in which interactions with others and rapid changes in technologies exert multidirectional, creative, and causal forces on writers' wayfinding to be a helpful terministic screen for focusing our own orientation to an emergent phenomenon.

However, we do also note that our own uptake of Alexander, Lunsford, and Whithaus' conceptualization of wayfinding is distinct from the group's development and application of the concept over a series of publications (Alexander et al., 2020; Lunsford et al., 2024; Whithaus et al., 2022). For the group, wayfinding began "as a metaphor for understanding" literacy and writing development (Lunsford et al., 2024, p. 273) that also became "a methodology for tracking how writers orient themselves or become oriented in multifaceted writing contexts" (p. 274). We, too, begin with wayfinding as a metaphor, or a terministic screen, for understanding and orienting to dynamism and complexity in emergent literacies and technologies. That metaphor then became a way of highlighting what Takayoshi et al. (2012) call a "*felt difficulty*" in interpreting and constructing our approach to this research (pp. 100-101), by helping us clarify that our goal is not to use social media rhetorics of prompt writing to uncritically inform or prescribe how students should (or should not) engage with prompt writing. Rather, our goal is to critically examine prompt writing rhetorics, invite students to do the same, and think with students about how to engage carefully and creatively with prompt writing as part of "a full quiver of semiotic modes" (Selfe, 2009, p. 645). And while, for Alexander, Lunsford, and Whithaus, the wayfinding metaphor led to a methodology that emphasizes phenomenological (Alexander et al., 2020) and contextual (Lunsford et al., 2024) understandings of writers' experiences in individual and group conversations, their wayfinding metaphor has guided us in taking an iterative and reflexive blending of computational and qualitative methods for examining social media rhetorics as a site for gauging dynamic and emergent literacy practices.

*2.2. Social media rhetorics scholarship in writing studies*

Scholars in writing studies and allied fields have long pointed to the impact that public discourses on literacies and technologies have on shaping students' and teachers' thinking about digital writing technologies and how studying those discourses can help support critical pedagogy initiatives (e.g., Gee, 2017; Gupta et al., 2024; Pigg, 2024; Vee, 2017). Computers and writing scholars have also long pointed towards the need to study "self-sponsored literacies" (Selfe and Hawshier, 2004, p. 232), particularly as a way of understanding how people engage with emerging and evolving technologies and media. X has been a useful platform for scholars in writing studies and allied fields interested in studying such "self-sponsored literacies" (Selfe & Hawshier, 2004, p. 232) or "way-finding" practices for navigating among writing contexts and engaging with newer media and technologies (Alexander et al., 2020, p. 124). This is why Buck (2023) argues that it is crucial to "understand how social media users integrate these sites within their daily literacy practices outside of the classroom" (p.17). Coad (2013) summarizes the position of many contemporary writing studies scholars about social media literacies well. Building on Johnson-Eilola and Selber (2009), he argues that while "the news media portray social networks as valueless forms of communication that are decaying young people's minds" (n.p.), social media literacies actually represent "sophisticated skills of understanding concrete rhetorical situations, analyzing audiences (and their goals and inclinations), and constructing concise, information-laden texts, as a part of a dynamic, unfolding, social process" (Johnson-Eilola & Selber, 2009, p.18 qtd. in Coad 2013, n.p.).

Supporting this idea, Moore et al. (2016) also point out in their survey of 1366 first year composition students in American universities that over 90.1 % students used Twitte for a range of literate practices. This sentiment is also reflected in Holmes and Lussos (2018) who analyze the "unique rhetorical affordances of Twitter bots as a way to offer student writers the kairotic means of understanding how networked writing functions in social media public spheres" (p.118). Coad (2017) himself has also studied the writing practices of graduate students on Twitter and found that far from being a haphazard activity, his participants used that platform strategically to extend their academic networks and to self-fashion professional identities for themselves. Buck (2023) also provides insights from her longitudinal case studies of various social media users highlights that her users developed the following key literacies through their social media use which are of vital interest to writing studies scholars: "(1) a heightened awareness of audience and an ability to tailor messages to specific audiences; (2) an understanding of how personal data is collected and circulated in online spaces; and (3) a means through which to use the first two skills for self-promotion and self-presentation in both personal and professional settings" (p. 8).

Though the platform has since changed, X has in the past been a useful archive for research in and beyond our field for several years now as it afforded free-flowing, interactive discussion in a micro-blogging format. Through its structuring of discussion using #hashtags, it has indexed historical moments based on popularity and allowed researchers to probe discursive trends that facilitate rhetorical actions, as demonstrated, for example, in our field's work on #BlackLivesMatter (Shelton, 2019), #OpKKK (Colton et al., 2017), #MyNYPD (Hayes, 2017), #GamerGate (Holmes & Lussos, 2018), #ArtAintFree (Amicucci, 2022), and #MeToo (Lang, 2019; Sarraf, 2023).

More recently, scholars have also been studying social media rhetorics to better understand the impact of generative AI technologies like ChatGPT. Pigg (2024) analysed YouTube videos of influencers using ChatGPT for research writing practices to build a descriptive taxonomy of emerging embodied literacy practices with AI. Englemann et al. (2024) also studied YouTube videos of introductory courses on artificial intelligence to argue the lack of significance given to ethics in many of them. Scholars like Li et al. (2023) have argued that a crowdsourcing approach with Twitter data on ChatGPT affords an opportunity to "tap into a diverse range of perspectives from the public" (p. 23). Like these scholars, we believe that social media rhetorics are themselves an important area of literate practice, and we also believe that social media rhetorics can function as a type of archive for conversations about writing with emerging technologies. Particularly because the researcher API for Twitter was still open at the time of our study, we approached social media data as a way to examine rhetorics about writing with and for ChatGPT. Even though the platform and access have since changed, which would require future methods to change accordingly, our engagement with social media data represents a snapshot in a moment of change–not only in emerging and expanding landscapes of generative AI engagement, but also in shifting landscapes of social media platform access and research engagement.

*2.3. Iterative, reflexive, dynamic methodologies for digital writing research*

Our iterative research process for examining social media rhetorics about generative AI follows a long tradition in computers and writing scholarship that recognizes that neither computers nor the methodologies and methods we use to study them are neutral; rather, computers, computer interfaces, and the methodologies and methods we use to study them are situated in relations of power and cultural contexts, and our positionalities in relation to them matter (Selfe & Selfe, 1994; Sullivan & Porter, 1997). Thus, as Sullivan and Porter (1997) model in framing "methodology as praxis" (p. 65), our approach was developed in recognition of the situatedness of knowledges and practices, and we embraced an iterative approach to our methods that was responsive to our positionalities and to changing dynamics in our context and in our relationship to our data. After all, as Van Kooten and Del Hierro (2022) argue, digital "methods and methodologies are learned, expanded, and understood best through experience" (p. 17), with particular attention to researcher positionality by a process of what Royster and Kirsch (2012) call "strategic contemplation," "a feminist orientation [that] asks us to 'pay attention to how lived experiences shape our perspectives as researchers'" (Royster & Kirsch, 2012, p. 22, qtd. in Van Kooten and Del Hierro 2022, p. 10). And as Riddick (2024) has modeled for social media research, specifically, a methodology of "deliberative drifting"—or "engaging with spontaneous, ephemeral activity on social media" (p. 40), requires not only being responsive to dynamic situations (in our case, emergent discourses about a newly launched tool on a platform that was itself undergoing changes), but also accounting for the ways our researcher positionalities and our data collection and processing choices are entwined (p. 50-51). We took inspiration from these iterative, reflexive, situated, and drifting methodologies partly because they align well with the "wayfinding" orientation to literacy (Alexander et al., 2020) that our work is grounded in, as it conceptualizes all meaning-making practices as dynamic, fluid, and constantly evolving.

*2.4. Computational methods blended with critical, rhetorical analysis in digital writing research*

We now turn to contextualizing machine learning as both the *what* and the *how* of our iterative analysis. Specifically, we acknowledge and build upon writing studies scholarship that engages with machine learning as both an object of examination and interrogation itself and as a method for analyzing large data sets. Highlighting the rhetorical nature of discourses and texts, scholars in our field have advocated that computational methods can become more rhetorically attuned through practices like augmenting textual data with contextual data, intervening at the level of training sets, and blending computational methods with critical, human-led methods. Indeed, Gallagher et al. (2020) argue that computational audience analysis methods should not be used to make "positivist, universalizing claims about the world," but should instead be understood as "deeply creative, generative, and messy endeavors" that cannot only interact with qualitative approaches, but also serve as an invention strategy by offering researchers a different scale to consider (p. 157).

For analyses using machine learning, scholars have also advocated for blending computational methods like topic modeling with critical and contextual analyses (e.g., Boettger and Ishizaki 2018, Hopton 2014, Kong et al. 2020, Lauer et al. 2018, Gray and Holmes 2022, Roundtree 2022, Tillery and Bloomfield 2022). For example, in Roundtree's (2022) analysis of codes of ethics for facial recognition technologies, "unsupervised topic modeling validated human coding by identifying and confirming reoccurring themes that emerged from human coding" (p. 213). In another study of a corpus of dissertations in rhetoric and composition, Miller (2022) blends qualitative coding and topic modeling. Writing with Licastro, Miller argues that "data doesn't speak for itself, but must be spoken into and from, based on deep disciplinary knowledge" (Licastro & Miller, 2021, p.8). Relatedly, scholars such as Mueller (2017), Tham (2023) use a mixture of distant and close reading, which is also a practice reflected in digital humanities scholarship like that of Underwood (2017).

Our own approach aligns with this thread of scholarship. To get a sense of how people have been orienting to and engaging with ChatGPT, which is trained through machine learning, we conduct a layered analysis of X discourses that began with distant reading (Gallagher et al., 2020) with machine learning methods and iteratively moved to closer, critical reading strategies. Given our commitment to critically examining social media rhetorics, we are especially inspired by Tillery and Bloomfield's (2022) layered approach to analyzing discourse in a climate skeptic Facebook group. Their analysis joins what they describe as distant reading methods (including topic modeling and a list of most frequently used terms) with a rhetorical analysis that involved a "close reading of the comments on [the Facebook group's] posts with the most interactions (i.e., likes and comments)" (p. 360). Tillery and Bloomfield describe the rhetorical analysis as a constellation approach that crucially emphasizes "argument patterns are not coincidental; they occur because they are linked to underlying ideologies and motivations" (p. 360). As a result, they argue, the "distant-reading methods offer a high-level view of the dataset, while the rhetorical analysis takes the findings and puts them into a meaningful context through the illustration of ideological constellations" (p. 361). We similarly join qualitative and qualitative readings of our data that attend to ideologies.

Our approach also aligns with the work of scholars who emphasize the material realities of computational research in our field. Gallagher and Beveridge (2022) advocate for a "project-oriented approach" to computational techniques that acknowledges the reality in our field that graduate programs do not typically front-load extensive training in computational methods and therefore "allows researchers to extend our field's technical and data-driven skill sets organically, based on the actual projects undertaken by researchers" (p. 12). Our own engagement with machine learning in this project is project-oriented and emerging, just as our analysis itself is emerging (and our relationship with ChatGPT is also emerging). And we also take inspiration from Graham and Hopkins' (2022) recommendations for researchers who use machine learning methods to provide accessible documentation of their methods in our own analyses, even as, following Buck and Ralston (2021), we provide only aggregated trends and do not provide the individual posts to respect the privacy of those whose posted.

## 3. Positionality statements

In taking up the tradition of reflexive, iterative approaches to digital writing and rhetoric research that we draw on, at the very outset, it is vital to account for researchers' positionalities, as they are part of the research ecologies that frame their work. To do this, we follow Jones et al.'s (2016) definition of positionality that accounts for the ways that identity markers are relational, historical, dynamic, and particular (p. 220), and we also follow Gonzales' (2022) call to "recognize the positionalit(ies) of the language(s) we use to do our work" (p. 98).

Anuj Gupta is a brown man from India, currently a graduate student in the U.S. who considers both Hindi and English to be his first languages, while also having limited conversational proficiency in Punjabi, Urdu, and Haryanvi. Ann Shivers-McNair identifies as a white woman from the U.S. whose first language is standardized white U.S. English. Both of us, as well as our research, is situated in our local context at a U.S. research university. We also acknowledge the assistance of ChatGPT in brainstorming and debugging code for our computational methods. Any coding suggestions given by ChatGPT were thoroughly reviewed and edited before use. We also acknowledge the resources provided by our university's high performance computing center to conduct our first run of topic modeling. In sharing our research findings as well as process for understanding emerging possibilities and practices at a pivotal moment in our relationship with AI, both of us hope to encourage our field to leverage its commitments to ethics in order to engage critically with rhetorics of writing with and for generative AI tools like ChatGPT.

## 4. Research methods

*4.1. Data collection*

To facilitate research, Twitter formerly offered researchers access to its API free of cost, using which researchers could download big datasets of posts as well as meta-data about them. We used this method to collect our data. Since then, X has changed its policies and most of its tiers of API access are now paid.

*4.1.1. Working with social media data*

Using a Python based library called *Twarc* (DocNow, 2023), guidelines from the social media platform's API, as well as Walsh's (2021) computational notebook, we created a Python script to download all posts published till May, 2023 that had the words "ChatGPT" and "writing" in them. We chose this search string to capture X users' articulations of how the former is impacting the latter. This process yielded a dataset of 258,661 posts. Following our IRB's requirements, we had a third-party broker de-identify our data to remove a bulk of the meta-data. The resultant dataset was collected in a comma separated values or .csv file which had posts with limited metadata.

*4.1.2. Ethics of X (formerly Twitter) data collection*

In working with social media, we needed to consider a range of ethical issues. First, we needed to check the platform's terms of use policies to ensure that we complied with its guidelines on how to access and use data. For example, sometimes websites can recommend download rates to prevent overloading their servers. In our case, since Twitter offered a dedicated API access to researchers when we collected data, we were able to adhere to their guidelines. Second, in terms of IRB compliance, "in most cases, this will not be considered human participant research, but caution is recommended before a researcher makes his/her own determination, because of the emerging ethical sensitivities in this area" (Walsh, 2021). We worked closely with our institution's IRB, who advised us to limit our data collection to the data types mentioned in Table 1 and then exempted us from human subjects research status for our study. Third, as researchers, we also needed to bring in our own ethics and consider the harms and benefits of the kinds of research we do. For example, as Buck and Ralston (2021), Walsh (2021) advise, posts should be thought of as existing on a spectrum of private and public, and while it is generally acceptable for social media researchers to present overarching trends on such posts, representing individual posts of common people without their consent can sometimes bring unwanted attention to individuals who might not want it. For these reasons, as well as the terms of our IRB, we share only aggregated trends and do not quote posts or provide the dataset we are analyzing in this research.

*4.2. Iterative data analysis*

Our research process, like the phenomenon we examine, is emergent, and therefore, we present our analysis as aligned with the iterative, reflexive spirit of methodology-as-praxis in computers and writing research (e.g., Sullivan and Porter 1997, Holmes and Verzosa Hurley 2024) and of project-oriented approaches to analytic tools and methods that Gallagher and Beveridge (2022) encourage. In the first few iterations, our analysis primarily involved understanding our data using descriptive statistics and then mapping out thematic patterns present in it using a technique called topic modeling. Based on these initial iterations, we found interesting subsets of data, and we gradually narrowed our focus to examine a subset of approximately 32,000 posts about 'prompts' and 'prompt writing'. To reach this stage, we used a range of computational, quantitative techniques using Python. To assist us in creating the code required for this analysis, we used both documentation provided by researchers who have used Python for similar kinds of analysis, assistance of ChatGPT itself in brainstorming and debugging code for our computational methods. Any coding suggestions given by ChatGPT were thoroughly reviewed and edited before use. We also acknowledge the resources provided by our university's high performance computing center to conduct our first run of topic modeling.

Once we obtained a smaller dataset focused on prompts, we used primarily human-driven qualitative coding to identify five themes in emerging rhetorics of prompts and prompt writing, and we analyzed those themes for implications for critical pedagogy and for further research. This decision to focus our analysis on prompts and prompt writing in our dataset happened in concomitance with emerging advisory guidelines by professional organizations in writing studies like the MLA-CCCC Joint Taskforce on Writing and AI about the vital need to teach students how to write ChatGPT prompts as a means to develop critical AI literacies (MLA-CCCC Joint Task Force, 2023).

**Table 1**
Metadata of our X (formerly Twitter) dataset.

| Column Name | Data type |
| --- | --- |
| text | Full text of the posts with a maximum character count of 240 characters |
| lang | The main language of a post as determined by X's internal algorithms |
| created_at | Date and time at which the post was posted |
| public_metrics.like_count | The number of likes a post received |
| public_metrics.quote_count | The number of times a post was quoted or reposted with additional commentary from the reposter |
| public_metrics.reply_count | The number of times users replied to a post |
| public_metrics.impression_count | The number of times a post was seen |

*4.2.1. Descriptive statistics*

In this section we provide an overview of our dataset using descriptive statistics to give our readers a general sense of the nature of our data. As discussed in the data collection section above, we have a dataset of 258,661 posts that contains their text as well as the following meta-data: language, date and time of publishing, public engagement metrics like retweets, likes, replies, and impressions. To understand the contours of this data, we used two libraries in Python. One is called pandas (McKinney, 2010) which helps analyze tabular data and the other is called matplotlib (Hunter, 2007) which helps to visualize that data.

We began by visualizing post distribution over time in our dataset, in order to contextualize emerging and shifting discourses. Fig. 1 presents the distribution of posts in our dataset roughly from late November 2022 to May 2023, visualizing the number of posts (from 0 to 10,000) on the y axis and dates on the x axis from November 2022 to May, 2023.

The two spikes visualized in the figure happened on 2023-02-08 (10,395 posts) and 2023-05-03 (5652 posts). On exploring posts on the first date, we did not find any particular news events that might have caused the spike. However, a majority of the posts on this day were retweeting posts written in the Thai language where the authors were sharing screenshots of their experiment with ChatGPT to create English sentences by feeding it bullet points of what they want to say in Thai and asking it to write a professional email in English with correct grammar and a polite tone. This post represents an important trend we found in our dataset: the discussion of strategies for multilingual speakers to use ChatGPT to create professional writing in English that sounded like first-language speakers.

The large amount of interest that users showed in this kind of discursive activity, which involved sharing guidance on best practices on how to write commands to ChatGPT for achieving functional literacy goals, also alerted us to the importance of the emerging discourse on prompts and prompt writing in our dataset. As we show in our analysis later, the themes we discovered in the X discourse on prompts and prompt writing reflect many ideas, concerns, and beliefs represented in what we first discovered in this initial spike in our larger dataset. This first spike also illustrates the diversity of languages in our dataset, even though we used English words as the search terms. As expected, English is the most widely represented language in this dataset, but there is also an important presence of other languages like Thai, Japanese, Korean, Chinese, Hindi, etc. We note this language diversity beyond the Anglosphere, even in an English language search, to highlight both the global nature of the discourse on ChatGPT and writing and the situatedness of our specific search process.

On the second date of a spike in posts, we found a significant news event. Most users were re-posting or talking about a news article published in Time magazine by Perrigo (2023b) that talked about the vote to unionize by 150 workers in Africa who worked as content moderators for AI powered companies like ChatGPT, Tiktok, and Facebook. Through this vote, Perrigo (2023b) reports, the first ever African Content Moderators Union was created. This represents another important trend we saw in our dataset: namely, the discussion of global flows of labor exploitation in the creation of AI technologies, as Perrigo (2023b) quotes a former ChatGPT content moderator:

> " 'For too long we, the workers powering the AI revolution, were treated as different and less than moderators,' said Richard Mathenge, a former ChatGPT content moderator who worked on the outsourcing company Sama's contract with OpenAI, which ended in 2022. 'Our work is just as important and it is also dangerous. We took an historic step today [in voting to unionize]. The way is long but we are determined to fight on so that people are not abused the way we were.'" (n.p.)

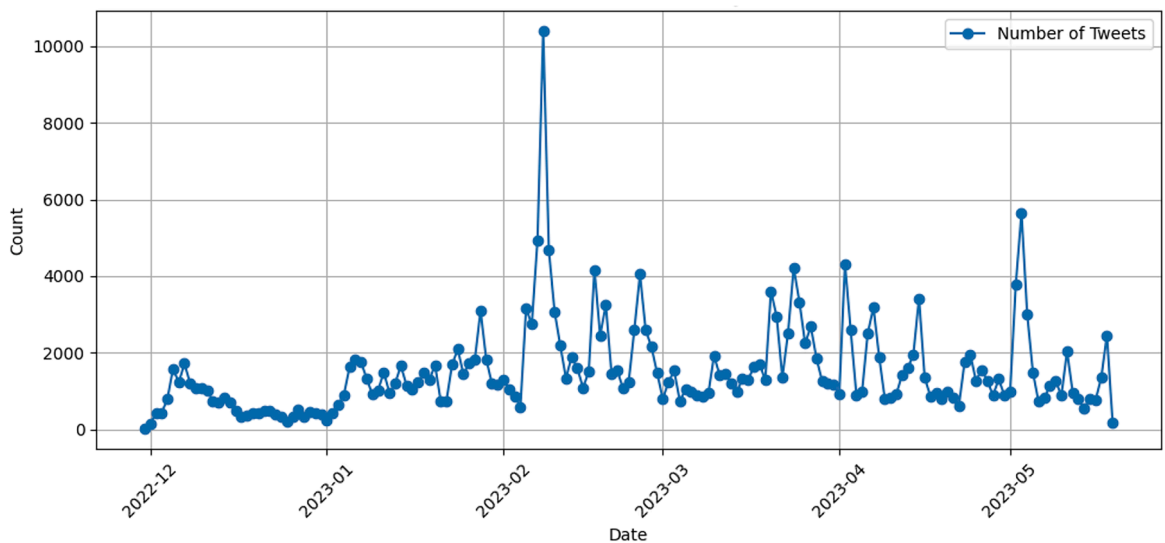

**Fig. 1.** Distribution of X posts over time.

The exploitation of Global South workers in ChatGPT content moderation is part of our research context and compels us to take a critical orientation, not only in our own engagement but also in the ways we talk about ChatGPT with students, to what Noble (2018) describes as "neocolonial trajectories" enabled by "neoliberal economic policies of globalization" that allow technology companies based in the Global North to outsource labor to low bidders that exploit Global South workers (p. 164). And as Roberts (2019) describes, such outsourced and exploitative labor is not limited to hardware and materials, but also includes content moderation for social media networks, software, and artificial intelligence models. *Time Magazine* reported, in the case of OpenAI's work on ChatGPT, that Kenyan-based workers were paid less than $2 per day to do the reinforcement learning work of labeling biased and harmful data in ChatGPT's training sets, which involved interacting for long periods with disturbing, violent, and toxic words and images (Perrigo, 2023a, 2023b).

Our engagements with these two spikes that we saw in our larger dataset shaped our focus in the study as part of our iterative research process. The first spike alerted us to the emerging literacy practice of prompting ChatGPT-like AI tools for achieving functional literacy goals and how it could potentially support social advancement for historically marginalized populations. The second spike helped remind us that the global effects of technology are always uneven and require a critical orientation to understand their large-scale socio-political impacts .

*4.2.2. Topic modeling*

Given what we learned from descriptive statistics, we then turned to topic modeling. While the two horizontal spikes that we learned about through descriptive statistics oriented us to the two topics of discussion in our dataset that caused the most engagement on single days, we also wanted to get a sense of the different topics that were being discussed across time. This is why we chose to use topic modeling as our next method for exploring our data. Topic modeling is a text analytics technique that reveals underlying patterns in a dataset, and it has been taken up across areas of writing studies (Gallagher et al, 2023; Graham & Hopkins 2022; Lauer et al. 2018; Majdik & Wynn, 2023; Messina, 2019; Omizo 2020; Tillery & Bloomfield 2022; Zhu et al. 2021; Roundtree, 2022). Within the tradition of machine-supported analysis of textual datasets, topic modeling presents a way to deal with limitations of older methods like keyword frequencies and collocations.

We can think of topic modeling as a computationally supported form of qualitative coding, i.e. as "an automated procedure for coding the content of a corpus of texts (including very large corpora) into a set of substantively meaningful coding categories called topics" (Mohr & Bogdanov, 2013, p. 546). Specifically, using algorithms like BERTopic (Grootendorst, 2022), topic modeling works "to reveal the latent structures of texts by discovering statistically salient clusters of words in texts based on the collocational frequency of words (what words usually go with other words)" (Gallagher et al., 2023, p.4). Collocations or measures to find the most statistically consistent words that accompany a keyword of interest, historically emerged in several disciplines like corpus linguistics as a means to capture semantic context. Topic modeling, then, is a more robust way to capture the meanings that collocations present by abstracting collocations into "topics."

Essentially, a topic is a constellation of collocations that recur in a manner that is statistically significant and thus signifies a potential topic or theme in the dataset. As Omizo (2020) puts it, "If the list of words per topic consistently co-occur, then they likely resonate with each other semantically. If so, then analysts can create categorical labels that explain this resonance" (p.3). The reasoning behind topic modeling thus is that "word associations signify the content of texts and that variations in word associations signify different content, which divide texts into coherent regions of meaning that humans easily understand as conversation shifts, argumentative development, or topics" (Omizo, 2020, p. 2-3). An example from Omizo (2020) will help to illustrate this: "A human reader might presume that the words 'gene,' 'dna,' and 'genetic' reflect a topic named 'genetics.' If this is so, then these words likely will cluster together across documents in contrast to other distributions, forming topics" (p. 3). Topics are not completely distinct, though: "topics created in the topic modeling process cross-pollinate; each topic in the topic model contains other topics, and enough topic models will contain multitudes" (Omizo, 2020, p. 6), which is why topic modeling is not an exact science. It requires interpretive interventions by human researchers based on their contextual and disciplinary expertise to help make sense of results through the categorical labels they assign to co-occurring words as well as through closer reading of the data to find connections between topics

Following the example of other scholars discussed above, we approached this computational approach as one step that would be further contextualized with a critically informed close reading. Omizo (2020) for example, helps us understand that "The topics of topic modeling are not simply themes; they might also reflect rhetorical frames, cognitive schemata, or specialized idioms" (Goldstone and Underwood, 2014, p. 361 qtd. in Omizo 2020, p. 4)." It can thus be productive, Omizo (2020) argues, to think of them using the rhetorical concept of "topoi" which provides three insights. First, it helps us see topics as heuristics or "stockrooms of received knowledge and as mobilizations of these common ideas by both rhetors and audiences" (p. 6). Second, "the social metabolization of topoi produces premises based on shared understandings" (p. 6). For example, "The common topoi of cause effect, for example, is characterized by predication because rhetors need to articulate the necessary relationship between one phenomenon and a prior phenomenon" (p.7). Third, topoi can be distinguished based on common and special topoi, with common topoi being "generalized, inferential rules for creating premises and special topoi as the material substantiation of these inferred premises that are conditioned by disciplinary domains, context, and audience interactions. Thus, a generalized "common topoi" such as "cause and effect" may apply to many different situations because it appears ontologically-fundamental or common" (p. 7). Following this topoi orientation, instead of using topic modeling to produce conclusive "topics" about our data, we used it as an exploratory technique to illuminate what patterns exist so that we could decide which pattern to dive deeper into with our own qualitative analysis.

There are several statistical techniques and Python based packages that one can use to do topic modeling. We used a package called BERTopic (Grootendorst, 2022). We chose BERTopic because it is one of the topic modeling algorithms that offer support for multilingual datasets. In operationalizing BERTopic, users can choose either language = "English", or language = "multilingual". If one chooses the "multilingual" parameter, then BERTopic is able to detect and model textual data from more than 50 languages. Therefore, while we acknowledge that the fact that our terms for querying the data using English keywords ("ChatGPT" and "writing") skews our dataset toward English language results (in other words, it only pulls posts from other languages that also include the English words "ChatGPT" and "writing"), we also emphasize the importance of mitigating uncritical defaults to English language information in big data research by highlighting the potential, however imperfect, of using a computational method that supports multilingual datasets.

Topic modeling algorithms develop two kinds of results. First, they produce a set of topics which are a collection of keywords that are often present together and thus are taken to mean a consistent topic. In other words, a topic modeling algorithm first tells us what it finds to be a collection of topics in the dataset. For each topic, it gives a list of keywords and quantifies the importance of each keyword for the topic. However, it cannot tell us what the topic means. It is up to the researchers to use their domain knowledge and close-reading of the data to interpret what a topic could mean. Secondly, topic modeling algorithms also tell us which topics are most likely to be present in each of the texts (X posts in our case) in the dataset. Thus, for each potential topic, researchers can see representative keywords as well as representative texts that make it up. They can then decide what a topic means based on their close-reading of both and can then start to infer relationships between topics to develop larger themes, as one usually does in qualitative coding methodologies like those popularized by Saldaña (2016). It is also important to note that topic modeling is an iterative process with researchers fine-tuning parameters as the interpretation process deepens. Many times, the algorithm creates topics that statistically make sense to it, but in terms of semantic and contextual meanings, they might be nonsensical and thus need to be filtered out by human readers.

Before running our topic model, we first cleaned the data to allow for more interpretable topics to emerge. Using a kind of coding technique called 'regular expressions' (Lopez and Romero, 2014) that is useful for cleaning textual datasets in python, we first converted the text of all the posts to lowercase, and then we removed all hyperlinks in them. Next we removed all punctuations and special characters so that only words would remain in our dataset. Once this data cleaning was complete, we ran our topic model. This initial running of the topic model using BERTopic (Grootendorst, 2022) yielded preliminary results in the form of 710 potential topics spread over 66, 965 posts. If the default settings of the algorithm are not able to cluster any texts in a dataset into a statistically significant topic, the algorithm does not include it in the analysis to avoid creating topics that are extremely small. Thus while our entire dataset contained 258,661 posts, out of those, BERTopic selected 66,965 posts as candidates for finding topics using its underlying statistical algorithm. While there are computational techniques through which various default settings of a topic modeling algorithm can be tweaked to force it to increase the number of texts it includes in its topics, we decided not to use those because the default settings are created by developers after extensive testing and validation and are thus more reliable. Tweaking them to force an algorithm to fit a dataset can also cause 'overfitting' or a scenario where highly idiosyncratic results are created that are not very interpretable and might not reflect broader patterns. Such limitations of computational methods are also an important reason why techniques like topic modeling might be helpful for providing initial exploratory insights rather than providing conclusive or objective analysis.

These initial results of 710 topics spread over 66,965 posts were available to us in the form of a Pandas dataframe (similar to tabular .csv data) which contained the following columns below in Table 2.

We share below the five most frequent of these 710 topics in Table 3.

As must be clear to our readers on looking at the representative words in Table 3, topic 1 relates to computer coding, topic 3 to journalism, topic 4 to tweets (now known as posts), and topic 5 to prompts and prompt writing. Topic 2 is a little more difficult to intuitively interpret. At this stage, we explored many of these 710 topics by closely reading the representative posts that made them up. Given how our descriptive analysis discussed earlier as well as the academic organizations in our surroundings were both pointing towards the importance of topic #5, i.e., the topic relating to prompts and prompt writing (MLA-CCCC Joint Task Force, 2023), we decided to focus the next iteration of our analysis towards this theme. This decision was also influenced by the fact that many of the other topics like topic #1 coding and topic #3 journalism also had extensive references to the impact of different styles of prompting ChatGPT on these areas of communication

### 4.2.3. Further data gathering

While the topic modeling algorithm had shared 627 posts that potentially represented the theme of prompts and prompt writing, we knew from our descriptive analysis as well as our own close reading of the data, that there were many more posts that were at least alluding to the concept of prompts. To cast a wider net and gather more of that data, next we decided to search for and create a grouping of posts from our larger dataset of 258,611 posts that contained the word "prompt." This led to a subset of 32,367 posts. In terms of methodological nuance, what this means is that while a larger number of posts discussed prompts in some way or the other, for the topic modeling algorithm only a smaller Section of them discussed prompts in a manner significant enough for it to include them as representative of the topic of prompts. For example, if a post contained more words that were representative of another topic like Topic 1: computer coding, and less words that were representative of Topic 5: prompts and and prompt writing, the algorithm would potentially annotate that post as representative of Topic 1 rather than Topic 5. Instead of going only by the annotations of the topic model, we decided to search for any post that might have even a passing mention of the word "prompt" for the next stage of our iterative analysis so that we could gather a more granular dataset about this topic or theme.

**Table 2**
Metadata from our topic modeling results.

| Column Name | Data type |
| --- | --- |
| Topic | A number alloted to a topic based on its frequency in the data. Topic 1 would have the highest frequency while Topic 710 would have the lowest. |
| Count | The number of posts in the dataset that contained the corresponding topic. In our cases this resulted in a total of 66,965 posts. |
| Representation | A list of 10 most frequent words that make up a topic. |
| Representative_docs | A sample text or posts in our case that the algorithm determines to have a high probability of containing the corresponding Topic. |

**Table 3**
Top 5 most frequent topics generated through topic modeling.

| Topic | Count | Representative Tokens |
| --- | --- | --- |
| 1 | 1232 | ['code', 'coding', 'copilot', 'github', 'codes', 'software', 'debugging', 'programming', 'developers', 'debug'] |
| 2 | 813 | ['quelle', 'guilds', 'rosesinthemath', 'クラスメソッド', 'vía', 'welp', 'awareness', 'bro', 'sorry', 'embrace'] |
| 3 | 800 | ['articles', 'journalism', 'news', 'article', 'headlines', 'journalists', 'journalist', 'headline', 'truth', 'reporter'] |
| 4 | 730 | ['tweets', 'twitter', 'tweet', 'threads', 'tweeting', 'thread', 'account', 'followers', 'viral', 'my'] |
| 5 | 627 | ['prompts', 'prompt', 'tips', 'effective', 'results', 'promptengineering', 'best', 'promptpal', 'better', 'examples'] |

Using another round of descriptive data visualization where we plotted the number of posts per day that contained the word "prompt" in our dataset (see Fig. 2), we learned that the number of posts about prompts in our dataset were gradually increasing over time. This further confirmed our assumption about prompts growingly becoming an important topic of discussion in posts about "ChatGPT" and "writing."

Thus we found that while quantitatively oriented techniques like descriptive statistics and topic modeling are helpful for honing into topics or themes of interest in a large dataset for initial exploratory analysis, it is helpful for researchers to complement those findings with their own intuition and reflexive understanding of their data as they move closer to analyzing some of those topics in more granular, qualitative detail. When used together in an iterative and reflexive manner, both kinds of methods can together can help researchers develop a more robust understanding of their data.

*4.2.4. Qualitative analysis*

In this section, we describe our iterative and collaborative process for qualitative analysis. To get a more granular look at how people were talking about the emerging phenomenon of "prompts," we started reading through our dataset of 32,367 posts. Since this dataset contained unique posts as well their reposts, we decided to further narrow our analysis to the 100 top most influential unique posts in this dataset. To decide which of these we should consider as the most influential, we decided to go by their "impressions" count which is a composite construct that X allocates to each post using a mixture of other parameters like replies, likes, re-posts, and quotes. Our intuition to use the impressions metric turned out to be correct as we found that these 100 posts represented the bulk of this dataset. Together, these 100 posts had been re-posted 26,507 times, making them approximately 82 % of our whole set of 32,367 posts that contain the word "prompt."

We then qualitatively coded these posts to see what kinds of patterns were emerging in how people are talking about prompts and prompt writing on X. Our approach to qualitative analysis was informed by Smagorinsky (2008). Counter to positivist approaches to qualitative coding, Smagorinsky (2008) advocates for a more constructivist one whereby "coding establishes the researcher's subjectivity in relation to the data and the framework through which data are interpreted. From this perspective, the codes are not static or hegemonic but rather serve to explicate the stance and interpretive approach that the researcher brings to the data" (p. 399). In this approach, researchers do not seek to develop inter-rater reliability in the conventional sense where two researchers work in isolation and then calculate statistical measures like Cohen's Kappa to ensure that their work is reliable. Rather, they labor together to "reach agreement on each code through collaborative discussion rather than independent corroboration" (Smagorinsky, 2008, p. 401). Through such collaborative coding, different researchers can bring their varying expertise and positionalities into the mix. This flexible approach can thus generate insights "because each decision is the result of a serious and thoughtful exchange about what to call each and every data segment" (Smagorinsky, 2008, p. 402). We viewed this approach as aligned with the dynamic and reflexive approaches called for in digital writing and social media methodologies by scholars like Riddick (2024).

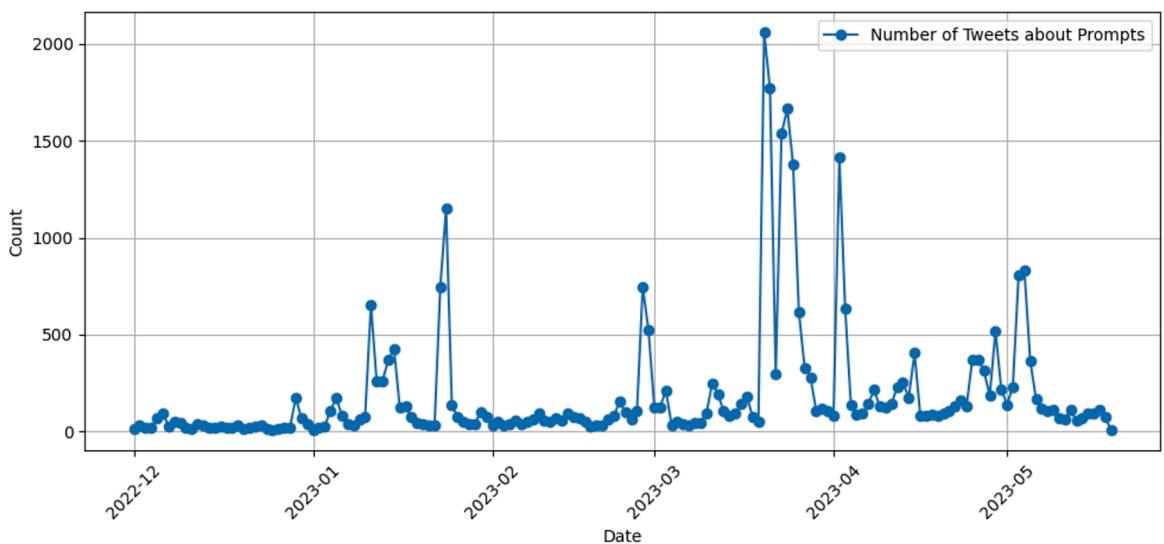

**Fig. 2.** Distribution of X posts that contained mentions of the word "prompt" over time.

To operationalize Smagorinsky's (2008) collaborative approach, we randomly divided our dataset of 100 posts into two sets of 50 each. Both of us first independently coded one of these datasets. While giving these initial codes, we followed Saldaña's (2016) definition of "initial coding." In this kind of coding, researchers first break down data into discrete parts, like we had done with our two datasets of 50 posts each and then closely examine the parts for similarities and differences (p. 115). Before annotating any codes, we read through the data to see what kind of patterns were reflected in them. With a goal of understanding an emerging writing technology and emerging literate practice that students are engaging with, as scholars like Alexander et al. (2020), Tham et al. (2021) call for, we focused our annotations at this stage on *in-vivo* highlightings of key phrases in the posts that illuminated the practice of prompt writing for ChatGPT. We also wrote memos reflecting on what those *in-vivo* phrases meant to us. Then we met to discuss what both of us had found.

Through this initial discussion, and given both of our grounding in disciplines of rhetoric, composition, and technical communication, we agreed that both of us were recognizing mini literacy narratives in our data. In these mini literacy narratives, X users were reflecting on their experiences of writing prompts with ChatGPT and sharing their emerging understandings of these human-computer interaction practices with other X users. We noticed certain common concepts or themes in these reflections. For example, both of us noticed how users were talking about prompts for ChatGPT in ways that also highlighted the potential impacts of ChatGPT on certain forms of communication that were relevant to them like emails, coding, copywriting, etc. Gupta noticed how users were also reflecting on specific formal or textual characteristics of prompting that worked or didn't work and many of them were also articulating their emerging definitions of prompting. Shivers-McNair noticed the presence of what we describe as micro-literacy resources like blogs and guidebooks on prompting that some of the posts shared along with a presence of advertising or marketing rhetoric to promote certain approaches to prompting.

In this stage of the discussion, we identified five categories, or larger codes, that we could use to annotate our data in more detail. As we discuss further in the next section, these included areas of communication impacted by prompt writing, micro-literacy resources for prompt writing, market rhetoric shaping prompting, rhetorical characteristics of prompts and definitions of prompts and prompt writing. We then went back to the data and started adding sub-codes for each post for each of these five codes. While creating these sub-codes, we were loosely following Saldaña (2016) definition of "axial coding." In this type of coding, researchers strive to create sub-codes that describe a larger code or category's "properties and dimensions" and how they relate to each other (p. 236). Upon completing this step, we once again discussed each other's sub-codes to decide on what we mutually agreed upon. In this discussion, we finalized sub-codes for each of our five main codes so that we could annotate each post accordingly.

## 5. Results and discussion

The results of our analysis are grounded in our iterative and collaborative process and in our orientation to prompt writing as both emergent and rooted in existing practices. As Warzel (2023) puts it, "Like writing and coding before it, prompt writing is an emergent form of thinking. It lies somewhere between conversation and query, between programming and prose. It is the one part of this fast-changing, uncertain future that feels distinctly human" (n.p.)–and thus, we add, an important area for writing researchers, teachers, and practitioners to engage. While we agree with Warzel that the specific configuration of human-machine relations in the current practice of prompt writing is emergent, we also acknowledge that the practice has many precedents and roots.

Indeed, the word "prompt" evokes a long history and rich current practice in rhetoric and writing studies of crafting carefully designed assignment guidelines for students, as showcased, for example, in the work of *Prompt: A Journal of Academic Writing Assignments*, which publishes innovative assignments contextualized with reflective essays. We also recognize prompt writing as a form of writing for algorithmic audiences (Gallagher, 2019), a rhetorical-technological practice that, as Haas (2007) reminds us, has precedents in Indigenous wampum practices that far predate the Western invention of hypertext. And like Koenig (2020), we recognize that writing for contemporary algorithms involves recognizing an interconnectedness of humans and algorithmic machines (p. 6). Given our goals of critically engaging students in conversations and practices of writing for generative AI tools, this lineage of "prompting" informs our orientation to the themes we identified in our data. In tracing this lineage of "prompting," we also recognize a dynamism and emergence that resonates with Alexander et al.'s (2020) conceptualization of wayfinding, which includes highlighting "the potential transience of the contexts in which people write" and focusing on "participants' fluid ability to not only move among those contexts, but also to create their own niches" (p. 124). Indeed, we recognize niche-creation moves across several of the themes we identified in our data. In what follows, we discuss five emerging themes from our qualitative analysis: (1) areas of communication impacted by prompt writing, (2) micro-literacy resources shared for prompt writing, (3) market rhetoric shaping prompt writing, (4) rhetorical characteristics of prompts, and (5) definitions of prompt writing.

*5.1. Theme 1: areas of communication impacted by prompt writing*

In our subset of the top 100 "prompt" posts, 58 % were assigned the code "areas of communication." This code refers to how posts spoke about the impact of prompt writing on existing areas of communication. We noticed a wide range of communication areas represented in the posts, so within this code, we applied the following (overlapping) sub-codes:

- 21 % of the posts refer to marketing (SEO, ad copy, content writing, tweets, blogs, newsletters)
- 17 % of the posts refer to academic writing (articles, summaries, student writing)
- 5 % of the posts refer to professional communication (emails)
- 6 % of the posts refer to creative writing (stories, books, publishing)
- 4 % of the posts refer to computer coding (apps, code, design specs)
- 2 % of the posts refer to personal writing (well-being journaling, personal growth strategy)

The fact that marketing communication is the largest sub-code is inextricable from the algorithmic audience strategies (Gallagher, 2019) deployed in many of the top 100 "prompt" posts (which we discuss further in the next two sections). In other words, it is not a coincidence that a focus on marketing strategies is present in a group of posts that accumulated the most impressions, likes, re-posts, and replies in our dataset. And while marketing communication is the most frequently referenced area, we believe it is also important to highlight the wide range of areas of communication impacted by prompt writing that we saw in our dataset, including practices that are not just for professional purposes. The fact that there were also many references to creative and personal writing practices in our dataset underscores Alexander et al.'s (2020) wayfinding orientation to understanding dynamic and mediated writing practices across contexts and domains, including "writing for purposes in addition to those defined by professional employment or roles within a professional activity system" (p. 123). Furthermore, the range of areas reflected resonates with our own observations and experiences of discourses connecting ChatGPT and other generative AI tools to a wide variety of writing tasks, with the potential to both disrupt or transform those practices, or to optimize or automate existing practices. This is why, when we teach about generative AI in our classrooms, especially when we discuss prompting practices, it is important to address how they impact not just academic writing or professional writing in isolation, but rather how they impact the wide range of transient literate practices that our students will engage in throughout their daily lives.

*5.2. Theme 2: micro-literacy resources shared for prompt writing*

We found that 68 % of the top 100 "prompt" posts refer to what we describe as micro-literacy resources for prompt writing. By this we mean bite-sized, modular, digestible resources for writing prompts, often shared via a link or directly in the post or post thread. We use the phrase "micro-literacy" to highlight the nature of X discourse itself, particularly an impressions-based attention economy. Within the micro-literacy resources code, we noticed a variety of resources, which we categorized with the following (overlapping) sub codes:

- 53 % of the posts referred to a prompt database (i.e. they included a list of prompts in the post or linked out to a larger database)
- 24 % of the posts referred to process knowledge (rules for writing)
- 3 % of the posts referred to a "master class" resource (i.e., a 60 s master class on prompts,)
- 10 % of the posts referred to tips and tricks
- 11 % of the posts offered resources in the form of tweetorials (Graham, 2021)
- 2 % of the posts referred to automatic prompt generators

The majority of micro-literacy resources offered were forms of prompt databases, i.e., a collection of ChatGPT prompts designed for specific outputs that users can copy or modify in their own interactions with ChatGPT. While, as we noted in our earlier discussion situating prompt writing in our field's perspectives, there are certainly connections to our field's practices of curating writing assignments, the distribution of knowledge across humans and machines in prompt writing (and prompt databases) connects perhaps even more precisely to what Hocutt et al. (2022) refer to as the microcontent paradigm in technical communication.

Hocutt et al. (2022) explain that "the microcontent paradigm extends component content management to contextualized and increasingly fragmented content delivery because of increased search-engine use across mobile and connected devices," noting that microcontent includes UX copywriting, or the text embedded in interfaces that guides users on how to interact with those interfaces (p. 117). In prompt writing databases, we see a similarly contextualized and fragmented approach not only to content delivery (i.e., the user-facing text that might be delivered via a chatbot or interface microcopy) but also to content production via pre-designed prompts for ChatGPT outputs. Hocutt et al. emphasize the need for skillful and ethical engagements with "content ecologies," which involve the "complex relations" of "user, content, metrics, and AI" (p. 127). We agree, and we believe that prompt databases are an increasingly important part of these content ecologies and, thus, of what Selfe (2009) has called the "full quiver of semiotic modes" (p. 645) from which students can select and that we, as teachers, can model critical and flexible ways of engaging.

As students navigate dynamic literacy practices related to genAI across domains, they are likely to be exposed to such micro-content pedagogic resources that are both usable, easily digestible, and often tinted with the ideological impulses of the marketing rhetoric that engulfs them. When we design resources in our classrooms, recognising this content should help us design both more usable as well as more critical lessons for our students. Furthermore, through the lens of Alexander et al.'s (2020) wayfinding, we also note a connection between the emergent form of prompt databases and writers' dynamic literacy practices that include "not only mov[ing] among … contexts, but also [creating] their own niches" (p. 124). In addition to critically adapting prompt databases, students can also position themselves as critical niche-makers in this emergent content ecology.

*5.3. Theme 3: market rhetoric shaping prompt writing*

We found that 51 % of the top 100 "prompt" posts engaged in what we describe as market rhetoric shaping the practice of and discourse about prompt writing. By market rhetoric, we mean the ways market forces are framing the value of prompt writing practices and resources. Within the market rhetoric code, we found the following (overlapping) sub-codes:

- 13 % of posts deployed a combination of deficit rhetoric (i.e., most people are bad at this) and resolution rhetoric (i.e., here is the solution)
- 15 % of posts deployed engagement transactions (i.e., sharing a resource for "free" in exchange for likes, retweets, etc.) to promote a micro-literacy resource such as a prompt database (described above)
- 5 % of posts emphasized the financial value (i.e., new jobs, increased salaries) of prompt writing
- 10 % of posts referred to prompt writing in terms of "super human" rhetoric (i.e., unlock your brain, supercharge your prompts)
- 10 % of posts deployed productivity rhetoric (i.e., boost your potential) to describe prompt writing
- 4 % of posts referenced market competition among generative AI tools, or what we might call "bot wars" (e.g., ChatGPT vs. Bard)

Given our discussion in the previous theme of the importance of attending to the role of prompt databases in content ecologies (Hocutt et al., 2022), we also want to emphasize the particular implications of the commodification of prompts via the engagement transactions we noticed, such as a poster offering a prompt database to be direct-messaged for "free" to people who like, reply, or retweet the post. A reflexive orientation to digital technologies and research (Selfe & Selfe, 1994; Sullivan & Porter, 1997; Haas, 2007; Royster & Kirsch, 2012; Van Kooten & Del Hierro, 2022; Holmes & Verzosa Hurley, 2024) calls us to recognize that the commodification of prompts via social media engagement is not neutral; it is value-laden, as engagement metrics drive what we find (and do not find), which in turn influences what we do (and do not do). Selber (2024) has also pointed out that the use of the phrase "prompt engineering" to refer to the process of writing prompts implies a certain appropriation of this phenomenon by a discourse of engineering that represents prompts as "silver bullets" that are being sold on AI marketplaces as magical solutions that can automate all writing.

We also recognize parallels between the rhetoric we observed in posts about prompt writing that emphasizes the potential for more value or capacity via prompt writing skill acquisition and the rhetoric of literacy myth/crisis, particularly as it has been applied to writing technologies (Johnson, 2023). As Wysocki and Johnson-Eilola (1999) caution, "When we speak then of 'literacy' as though it were a basic, neutral, context-less set of skills, the word keeps us hoping—in the face of lives and arguments to the contrary—that there could be an easy cure for economic and social and political pain, that only a lack of literacy keeps people poor or oppressed" (p. 355). While the posts themselves do not tend to use the word literacy to describe prompt writing skills, we recognize a similar tendency to connect a de-contextualized skill with resource acquisition. We also recognize that, as writing teachers, our own impulse to describe prompt writing as a "literacy" (evidenced in the previous section) reflects not only the sedimentation of that term in our field despite Wysocki and Johnson-Eilola's (1999) cautions, which Tham et al. (2021) likewise acknowledge, but also the forces of market rhetoric that shape how we frame the value of what and how we teach, including and especially our engagement with generative AI. At the same time, connecting prompt writing with literacy also creates a pedagogical space for emphasizing a critical awareness of the "complex and recursive movement" across writing ecologies that Alexander et al.'s (2020) wayfinding literacy metaphor highlights. In this case, such an awareness involves recognizing a literacy practice that may be distributed not only across a human and the generative AI tool they are using, but also across multiple contributors to a prompt database used in a specific genre performance.

*5.4. Theme 4: rhetorical characteristics of prompts*

In the top 100 "prompt" posts, 21 % described rhetorical characteristics of prompts, by which we mean strategies associated with effective prompt writing. Within this group, we found the following (overlapping) sub-codes:

- 3 %: word count (e.g., claims like longer prompts are better)
- 1 %: multimodal prompts (e.g., using audio inputs to write prompts)
- 4 %: tone (e.g., customizing ChatGPT to your writing style)
- 1 %: template (e.g., using and customizing existing prompts)
- 3 %: data from other sources (e.g., incorporating samples of your writing to train ChatGPT)
- 3 %: correct/incorrect rhetoric (e.g., your prompt is incorrect!)
- 2 %: garbage in garbage out (GIGO) (e.g., the better the prompt, the better the output)
- 3 %: imperative statement (e.g., command ChatGPT to write an essay)
- 1 %: interrogative statement (e.g., asking ChatGPT questions)
- 3 %: evocative (e.g., using please and thank you when addressing ChatGPT)

We note, first, that even though these percentages are low frequencies, the posts in this subset have thousands of retweets in the larger dataset, and thus we believe these patterns are worth considering and investigating. We also note that the nature of our data and IRB stipulations requires us to focus only on the texts of posts, which in many cases included a link to an external resource with prompts without including the actual prompt(s). Still, in sample prompts and in commentary and advice about prompts that appears in post texts in our subset, we noticed that prompt writing practitioners seem to be shifting from interrogative queries (typical of many people's interactions with search engines) toward imperative commands for generative AI (with considerable, often humorous debate over whether or not we should be saying "please" and thank you"). In this shift we recognize a potential trend in emerging human-computer interaction practices worth tracking further over time, as prompt writing practices and prompts proliferate.

We also note a range of orientations to the still-emerging conventions of prompting. We note the influence of the GIGO ("garbage in, garbage out") mental model in the way many posts frame prompt writing, which we often associate with computer code, and which we see as related to people's perceptions of prompts as either "correct" or "incorrect." By contrast, we associate the attention, in some posts, to the importance of tone and style directions to a rhetorical perspective, where a prompt might not be incorrect or correct so much as more or less effective. This observation brings us back to Warzel's (2023) postulation that prompt writing "lies somewhere between conversation and query, between programming and prose." This "in-between-ness" appears to bear out in the discourses in our dataset and is worth tracking as prompt writing discourses and practices expand, because the "in-between-ness" of prompt writing may have profound implications for genre-specific knowledge and genre awareness on a rhetorical, social/contextual level in genre performances involving generative AI, as we discuss further in the conclusion.

While these techniques may not yet be typified, certain patterns are crystallizing. Mollick (2023), for example, has postulated two emerging styles of prompting. One, which he calls "conversational prompting," involves free-flowing conversations with ChatGPT like tools to receive desired outputs, and the other, which he calls "structured prompting," involves writing out long, structured prompts using templates to customize the kinds of outputs produced. In what we might characterize as a structured prompting approach that also attends to the rhetorical situation, Ranade et al. (2024) have developed a formula for a human-in-the-loop approach to rhetorical prompt writing: "Prompt = (*audience* + *genre* + *purpose* + *subject* + *context* + *exigence* + *writer*)" (n.p.). The goal of this template, Ranade et al. explain, is for writers to attend to "which audiences are excluded from the conversation" and whose subjectivities are centered in defining "audience needs" (n.p.). To return to Alexander et al.'s (2020) conceptualization of wayfinding that includes creating niches for literate practices as well as moving among contexts, we recognize in Ranade et al.'s approach a niche-making that can push prompting patterns to include not only attention to the forms of prompts, as evidenced in the sub-codes in our data, but also a more critical attunement to how prompts are designed for audiences.

*5.5. Theme 5: definitions of prompt writing*

Of the top 100 "prompt" posts, 15 % included definitions of prompt writing, by which we mean the ways people describe what prompt writing is and does. Within this group, we identified the following (overlapping) sub-codes:

- 10 %: usability (prompt writing as what end-users can do to make ChatGPT usable)
- 3 %: evaluation (prompt writing includes evaluating the outputs)
- 2 %: skill (prompt writing as a competency)
- 1 %: talking (prompt writing as a means of talking to machines)
- 1 %: thinking (prompt writing as an emergent way of thinking)
- 1 %: search engine (prompt writing as a new interface for searching the internet)

The emphasis in these posts on prompt writing as a way that end-users can make ChatGPT more usable evokes a connection to writing studies scholarship on localization, or the practice of contextualizing technologies and content for specific communities (e.g., Agboka 2013, Sun 2012). While some localization work happens on the developer side as designers apply insights from audience analysis or user experience research, scholars like Sun (2012) draw attention to user localization, which "emphasizes the contributions users have made to a technology's design process in a participatory culture" (p. 41). We connect this orientation to user localization practices with wayfinding literacy practices, especially the ways that people engage with emerging writing technologies, move among contexts, and create niches that Alexander et al. (2020) highlight. Therefore, the proliferation of prompt databases and prompt sharing on social media and other platforms is a vector through which local practices can reach global scale and then be re-localized in new contexts, across languages and cultures. Thus, it is particularly important to bring critical awareness to the process.

## 6. Conclusion

We now turn to a discussion of pedagogical and research implications and recommendations for engaging with prompt writing. Specifically, we highlight the ways in which a wayfinding-inspired pedagogy can help us, as teachers, support students in taking up prompt writing, particularly in academic and professional contexts. The themes we found in our dataset suggest that the wide ranging areas of communication impacted by prompt writing and the wide ranging rhetorical practices and values associated with prompt writing necessitate the kind of complexity, capaciousness, reflexivity, and dynamism that a wayfinding orientation to literacy (Alexander et al., 2020) affords. We also highlight directions for future research on prompt writing that "embrace multiliteracies beyond [those associated with] professional employment" (Alexander et al., 2020, p. 123).

*6.1. Connections to pedagogy*

Given these emerging rhetorics and their resonances with Alexander et al.'s (2020) wayfinding orientation to literacy, we suggest that instructors can engage students in a collaborative, critical engagement with prompt writing as an evolving literacy practice. We can invite students into the process of refining the emerging rhetorics of prompt writing we have identified at scale in a social media dataset in conversation not only with emerging scholarly and professional treatments of prompt writing, such as Ranade et al.'s (2024) rhetorical formula for prompt writing, but also with their own experiences and observations. Indeed, Alexander et al. take a phenomenological approach to engaging closely with the rich and contextualized experiences of individual student writers, arguing that "Part of being more attentive to lived experience means that wayfinding tracks an individual's agency in determining what, when, and how to write, as well as in defining what writing should be valued–and how that agency is complicated by discourses, objects, dispositions, and other elements that participants and researchers identify" (p. 124). While Alexander et al.'s aim is to theorize and conceptualize "diverse writing experiences" in our field's research (p. 126), we recognize in their approach the potential to invite students to reflexively investigate their own engagements with rhetorics of prompt writing and specific literate practices of prompt writing.

For example, to return to the micro-literacy theme (#2) we highlighted above, we suggest that creating, sharing, remixing and using prompt databases is an important site for guiding students in critically and reflexively tracking their own agency and distributed agencies in using generative AI across the multiple literate domains they inhabit. Such a wayfinding approach to agency tracking could also involve tracing layered ideologies and values embedded in tools like ChatGPT, in prompt templates and practices, and in specific contexts. For example, Gupta (2024b) has created a chatbot called *Translinguo* by repurposing some of these prompting techniques to support critical language awareness goals. While ChatGPT and other popular chatbots largely use standardized American English, *Translinguo* has been prompted by Gupta (2024b) to respond in a variety of World Englishes like Hinglish, Spanglish, Jamaican English etc. in order to help normalize a translingual view of language in a technical communication context. Using this model, students can be encouraged to use prompt writing to achieve not just functional but also critical, social justice goals.

To return to the theme (#1) of wide ranging areas of communication impacted by prompt writing, we believe that the increasing ubiquity and automation involved in wide-ranging ChatGPT applications across areas of communication makes it important to help students understand different levels of prompting and automation that might be appropriate in different contexts. Additionally, as students study the varying impact on different communication practices by generative AI tools, it also becomes all the more important to create space for students to attend to the inequities that are built into the ecology of the technology itself, such as the exploitation of Global South content moderators for ChatGPT and their activism in unionizing (Perrigo 2023a, 2023b). Additionally, this would also involve teaching students to critically examine rhetorics of prompt writing they encounter in their own spheres by paying attention to the ideological impulses and market rhetorics (theme #3) that underpin the micro-literacy resources for prompt writing, as well as definitions and practices of prompt writing (themes #4 and #5) that students access the digital spaces around them.

This wayfinding-inspired, critical-reflexive approach to situating students' engagements with an evolving technology and emerging rhetorics of prompt writing for that evolving technology could be paired with an application of Ranade et al.'s (2024) rhetorical approach to prompt writing in which students use the rhetorical prompting formula, "Prompt = (*audience* + *genre* + *purpose* + *subject* + *context* + *exigence* + *writer*)" (n.p.), to both analyze existing prompts and to generate and test their own prompts. Therefore, a combined critical-reflexive and applied approach might look like a sequence of activities that guides students through critically contextualizing first generative AI technologies themselves, then prompt writing rhetorics and practices. For example, such a sequence might include:

- Discussion of power, labor, and linguistic and cultural representation in the creation, curation, and use of generative AI technologies and in the creation, curations, and use of prompts
- Discussion of prompt writing and rhetorical agency across academic, professional, and public perspectives, including students' own engagements or investigations
- Modeled and collaborative selection, contextualization, and analysis of prompts (or a prompt database) using Ranade et al.'s (2024) rhetorical prompt writing formula
- Individual or group-based student projects that analyze or develop a prompt database for a specific rhetorical community and purpose, perhaps including feedback on or testing of prompts in the community

*6.2. Directions for further research*

We recognize that our own scholarly understanding of prompt writing, like the discourses and practices of prompt writing themselves, is emergent, and there is more work to be done. We also recognize that there is more work to be done on the question of social media research methods for studying discourses of prompt writing, as the particular social media platform featured in this study has since undergone API changes that prevent the data collection process we used. Such changes may be indicative of larger implications for social media research, as the Coalition for Independent Technology Research notes in the case of Reddit's API changes (Gilbert et al., 2023). Therefore, we hope our work can mark a snapshot-in-time of discourses, practices, and methods in a pivotal year for generative AI and for social media research that provides a point for future reference and further consideration in research, teaching, and practice. Indeed, Gupta is building on the current study in a larger project that takes a deeper dive into "prompts" and "prompt writing" to study the impact of generative AI tools on literacy through a mixed methods corpus driven approach (Carradini and Swarts, 2023). In this project, Gupta explores "prompts" as an emergent AI genre and "prompt writing" as an emergent AI literacy by a collecting a corpus of prompts for academic, technical, and business communication published in open-access writing studies journals to understand how scholars in writing studies are defining this emerging human-computer interaction practice, and what affordances and constraints they bring for the writing, rhetoric, and technical communication classroom.

We also hope that highlighting wide ranging and emergent rhetorics of prompt writing at scale can complement ongoing and future research that examines the lived experiences of writers who "embrace multiliteracies beyond professional employment," like participant Miguel in Alexander et al.'s (2020) study of wayfinding practices across writing contexts (p. 123). We agree with Alexander et al. that such phenomenological approaches are important for understanding "the potential transience of the contexts in which people write," as well as people's "fluid ability to not only move among those contexts, but also to create their own niches" (p. 124). And we believe that it is especially important to include and center traditionally underrepresented students' and writers' perspectives, insights, and expertise as we continue to learn more about what prompt writing looks like and how writers can engage critically and effectively in prompt writing. Our field is continually expanding our understanding of writing practices by expanding our attention not only to new and emerging technologies and sites of writing but also by attending to long-standing but historically marginalized technologies and sites of writing (e.g., Del Hierro, 2019), and our field is well positioned to continue doing this work in relation to prompt writing.

**CRediT authorship contribution statement**

**Anuj Gupta:** Writing – review & editing, Writing – original draft, Visualization, Validation, Resources, Project administration, Methodology, Investigation, Formal analysis, Data curation, Conceptualization. **Ann Shivers-McNair:** Writing – review & editing, Writing – original draft, Validation, Supervision, Resources, Project administration, Methodology, Investigation, Formal analysis, Data curation, Conceptualization.

**Declaration of interest and funding**

The authors have no competing interests or external funding to declare for this work.

**Data availability**

The authors do not have permission to share data.

Anuj Gupta is a doctoral candidate (ABD) in the Department of English at the University of Arizona. In the past, he has helped build one of India's first writing programs at Ashoka University. A winner of *Kairos'* Graduate Student Research Award & the AACU's K. Patricia Cross' Future Leaders Award, Anuj's research has appeared in refereed journals, edited collections, and proceedings. He is currently working on his dissertation where he is studying the impact of Generative AI chatbots on academic, technical, and business communication by analyzing ChatGPT prompts as an emerging genre of writing.

Ann Shivers-McNair is Associate Professor and Director of Professional and Technical Writing in the Department of English and Affiliate Faculty in the School of Information at the University of Arizona. She is the author of *Beyond the Makerspace: Making in Relational Rhetorics* (University of Michigan Press, 2021), and her research has also appeared in refereed journals, edited collections, and proceedings. Her collaborative work has received national recognition, including a National Science Foundation grant and the Society of Technical Communication Frank R. Smith Distinguished Article Award.